\documentclass{iopjournal}
\usepackage[export]{adjustbox}
\usepackage{graphicx, subcaption} % Required for inserting images
\usepackage{orcidlink}

\usepackage{amssymb}
\usepackage{amsmath}
\usepackage{algorithm}
\usepackage{algpseudocode}
\usepackage{bm}
\usepackage{hyperref}
 
\usepackage{placeins}

\usepackage[
backend=biber,
style=numeric-comp, %style=numeric
sorting=none
]{biblatex}
\begin{document}

\articletype{Paper} %	 e.g. Paper, Letter, Topical Review...

\title{Towards Real-time Control of a CartPole System on a Quantum Computer}

% Approaching Real-time Control of a Cartpole System on a Quantum computer

% Decision on authorship (09.12.)
% Seth & Väinö

% 0009-0001-2350-7512
% Jerome
% Peiyong )_
% 0000-0002-6359-8922 Nguyen Truong Thu Ngo (nguyentruongthu.ngo@adelaide.edu.au)

% 0009-0008-7933-6848 Jerome Lenssen (jerome.lenssen@vtt.fi)
% Francesco Cosco - 0000-0002-4418-3879
% James Quach - james.quach@csiro.au - 0000-0002-3619-2505

\author{Nguyen Truong Thu Ngo$^{1,2,\dagger, *}$\orcidlink{0000-0002-6359-8922}, Väinö Mehtola$^{3,\dagger,*}$\orcidlink{0009-0001-2350-7512}, Jér\^{o}me Lenssen$^{3}$\orcidlink{0009-0008-7933-6848}, Peiyong Wang$^{2,*}$\orcidlink{0000-0002-0665-6639}, Francesco Cosco$^{3}$\orcidlink{0000-0002-4418-3879}, Tien-Fu Lu$^{1}$\orcidlink{0000-0001-9757-9028},  and James Q. Quach$^{2,*}$\orcidlink{0000-0002-3619-2505}}

\affil{$^1$School of Electrical and Mechanical Engineering, Adelaide University, Adelaide, Australia}

\affil{$^2$Commonwealth Scientific and Industrial Research Organisation (CSIRO), Clayton, Victoria 3168, Australia}

\affil{$^3$VTT Technical Research Centre, Espoo, Finland}

\affil{$^\dagger$These authors contributed equally and share first authorship.} \\
\affil{$^*$Author to whom any correspondence should be addressed.}

\email{nguyentruongthu.ngo@adelaide.edu.au, vaino.mehtola@vtt.fi, james.quach@csiro.au}

\keywords{quantum reinforcement learning, robotics, quantum machine learning, NISQ, real-time control}

\begin{abstract}
The application of quantum reinforcement learning (QRL) to real-time control systems faces significant challenges regarding hardware latency, noise susceptibility,  and learning convergence. This work presents an end-to-end investigation of a minimal hybrid quantum-classical agent applied to the CartPole benchmark, addressing the gap between idealized simulation and execution on a physical superconducting quantum processing unit (QPU). We demonstrate that a single-qubit agent acts as an effective learning model, solving the environment in substantially fewer episodes than a comparable classical actor-critic network even when the training of the hybrid agent is restricted to use parameter-shift for its quantum circuit component. To connect learning to deployment constraints, we map the inference-time trade-off between control-loop rate and measurement shot budget to provide guidance for an eventual real-time control demonstration. The resulting performance matrices show that both inference control frequency and shot count strongly affect balancing stability: higher inference frequencies consistently improve performance, and increasing the shot budget lowers the minimum inference frequency required to achieve near-maximal balancing. These results highlight the importance of finding an optimal medium between shot count and control frequency and developing circuits that are e.g. initial-state invariant. Lastly, we address the critical bottleneck of control latency on NISQ hardware. By bypassing the standard high-level software stack and programming the Zurich Instruments readout electronics directly via command tables, we achieve more than an order-of-magnitude improvement in execution speed on the VTT Q5 processor. These results quantify some of the current boundaries of quantum-assisted control and provide a start for achieving the tens-of-hertz throughput required for real-time closed-loop control feedback.
\end{abstract}

\section{Introduction}

Quantum machine learning (QML) investigates how quantum information processing can be leveraged for learning tasks, motivated by the fact that quantum systems can naturally represent and manipulate high-dimensional structures that are classically intractable~\cite{schuld2015introduction,mcclean2016theory,biamonte2017quantum}. 
From a theoretical perspective, several works have identified learning settings in which quantum models admit provable advantages over classical approaches, typically under complexity-theoretic or cryptographic assumptions~\cite{biamonte2017quantum,liu2021rigorous,gyurik2023exponential}. 
While these results establish an important conceptual foundation for QML, they are often derived in idealized regimes and do not directly translate to near-term hardware nor practical applications. 
As a result, practical QML research has largely focused on hybrid algorithms compatible with noisy intermediate-scale quantum devices, where quantum circuits are embedded within classical optimization loops~\cite{preskill2018quantum,cerezo2021variational,callison2022hybrid}. 
This paradigm is most commonly realized through parameterized quantum circuits (PQCs), in which gate parameters are treated as tunable variables that can encode classical data and be optimized using classical gradient-based or gradient-free methods~\cite{benedetti2019parameterized,schuld2019quantum}.

Reinforcement learning (RL) is a general framework for learning sequential decision-making policies from interaction with an environment, and it has become a standard methodology for, e.g., control and robotics when accurate system models are unavailable or difficult to derive~\cite{hwangbo2019learning,tan2018sim,kaufmann2023champion,peng2020learning,kohl2004policy}. Beyond continuous-control benchmarks, RL has demonstrated high-profile successes in complex domains that require long-horizon planning and reliable generalization, including self-play systems that achieve superhuman performance in chess and the board game of shogi \cite{Silver2017-hj}, large-scale agents for real-time strategy video games such as StarCraft II \cite{Vinyals2019-at}, and learned controllers for challenging robotic locomotion tasks \cite{Peng2020-bt}. Despite its successes, RL methods can remain sample-inefficient and exhibit sensitive training dynamics as problem dimensionality grows, and policies may lose performance when deployed under conditions that differ from those encountered during training.

Quantum reinforcement learning (QRL), a subfield of QML, applies quantum computing methods to reinforcement learning with the aim of improving aspects such as sample efficiency, exploration--exploitation trade-offs, and parameter count relative to fully classical approaches~\cite{dong2008quantum,dunjko2016quantum,saggio2021experimental,paparo2014quantum,hamann2021quantum,dong2010robust}. While there are no proven theoretical guarantees for QRL providing advantages over classical methods to date, prior work has focused on providing experimental results where QRL performs well in sample efficiency or parameter count \cite{saggio2021experimental,dong2008quantum,paparo2014quantum,hamann2021quantum,dong2010robust, chen2020variational,lockwood2020reinforcement,jerbi2021parametrized,skolik2022quantum,lan2021variational,chen2022variational,yun2022quantum}. For the algorithms themselves, while primitive-based QRL approaches have been studied for improving sample efficiency, such as in the Grover-based works in~\cite{saggio2021experimental,dong2008quantum,paparo2014quantum,hamann2021quantum,dong2010robust}, more general PQC-based QRL has been explored in the context of the near term quantum hardware constraints~\cite{chen2020variational,lockwood2020reinforcement,jerbi2021parametrized,skolik2022quantum,lan2021variational,chen2022variational,yun2022quantum}. 

As a representative benchmark for comparing classical and quantum reinforcement learning approaches, prior work has frequently considered the CartPole task, owing to its simplicity, low-dimensional state space, and suitability for assessing learning efficiency~\cite{Riedmiller2005,Geva1993,Lockwood2020RL,skolik2022quantum}. Within this setting, a growing body of quantum reinforcement learning (QRL) research has explored parameterized quantum circuits (PQCs) as trainable components of the policy and value function, primarily to study learning behavior, parameter efficiency, and feasibility on near-term quantum hardware.

Several earlier studies have investigated PQC-based agents on CartPole and related control tasks without an explicit focus on real-time deployment nor any hardware implementation results. Chen et al., 2020. \cite{chen2020variational} implemented a value-based approach where a PQC computes the state-action value (Q) function based on deep Q-learning \cite{mnih2013playing}. The agent would sample the action corresponding to the highest calculated Q-value at the quantum circuit output. Chen's model represented the earliest PQC model functioning as value function approximator, which was later extended in Jerbi et al's work as policy approximator \cite{jerbi2021parametrized}. Jerbi et al.~\cite{jerbi2021parametrized} introduced the SOFTMAX-PQC model, in which a softmax policy is derived from expectation values of a multi-qubit PQC with trainable observable weights, input scaling parameters, and variational gate angles. The proposed models, based on classically hard yet trainable PQCs, were evaluated on standard OpenAI Gym environments including CartPole-v1, MountainCar, and Acrobot using a Q-learning algorithm, but did not consistently meeting the solving conditions of the learning tasks. Skolik et al.~\cite{skolik2022quantum} investigated PQC-based deep Q-learning agents on OpenAI Gym benchmarks including CartPole, focusing on architectural design choices such as data encoding strategies, readout observables, and circuit depth, and demonstrated that learning performance is highly sensitive to these design decisions. Hsu, Lin, and Wang~\cite{hsu2025comparative} presented a comparative study of quantum and classical reinforcement learning on the CartPole task, using simulated parameterized quantum circuits trained with Proximal Policy Optimization and reporting competitive performance with fewer trainable parameters and training episodes relative to classical baselines. Collectively, these studies provide insight into the learning behavior and architectural considerations of PQC-based agents, but remain limited to simulation-based evaluation and do not address hardware execution constraints or real-time control considerations.
 \iffalse A elated study have also examined simplified pendulum balancing problems without cart dynamics. For example, Lan~\cite{lan2021variational} proposed a multi-qubit variational soft actor-critic algorithm and evaluated it on a single-degree-of-freedom pendulum balancing task, reporting reduced parameter counts relative to classical baselines while achieving comparable control performance. \fi

More recent studies have begun to address aspects of hardware execution that are relevant to real-time control. Hsiao et al.~\cite{Hsiao2022-uj} investigated unentangled single- and multi-qubit PQC architectures trained with Proximal Policy Optimization (PPO), demonstrating successful learning on CartPole, Acrobot, and LunarLander environments and executing the trained policies on an IBM quantum processor, thereby addressing practical aspects of hardware inference relevant to real-time control, albeit without a systematic latency or stability analysis.

Among existing studies, Sun et al.~\cite{sun2025first} are most closely aligned with the focus of this work. They trained a variational quantum circuit policy using a model-based offline reinforcement learning approach, in which a dynamics model is learned from a fixed dataset of pre-collected physical CartPole environment interactions and subsequently used for policy optimization, and demonstrated its deployment in a real-time CartPole control experiment. Notably, their implementation relied on an eight-qubit quantum policy executed on embedded hardware via a Raspberry Pi--based quantum simulator to partially control a physical CartPole system. They further analyzed the latency constraints of cloud-based quantum execution, reporting action latencies exceeding three seconds when using standard IBM QPU access~\cite{sun2025first}, which rendered such execution unsuitable for latency-sensitive control. Consequently, their results highlighted the need to rely on simulator-based execution for real-time operation. As a result, realizing closed-loop CartPole control directly on a physical quantum processing unit—where the agent must simultaneously contend with hardware noise and execution latency—remains an open challenge.

In this paper, we study PQC-based reinforcement learning on CartPole with an explicit focus on the gap between simulation-only evaluations and latency- and noise-constrained deployment. Our goal is to assess how a minimal hybrid quantum--classical agent behaves (i) during training under shot-based gradient estimation, (i(ii) under different inference-time constraints such as control-loop rate and shot budget, and (iii) when executed through a low-latency hardware control path that approaches real-time feedback requirements.

We adopt an actor--critic policy-gradient baseline and augment it with a single-qubit variational circuit (five gates) adapted from~\cite{Hsiao2022-uj} as the quantum component. We benchmark a fully classical agent against the hybrid agent on CartPole-v1~\cite{Brockman2016-lq}. A central contribution is that the hybrid agent retains its sample-efficiency advantage not only under analytical (expectation-value) gradients but also when trained using the parameter-shift rule with finite-shot circuit evaluations.

We then move from training to deployment constraints by mapping the inference-time trade-off between control-loop rate and sampling precision. Concretely, we evaluate how balancing performance changes as the inference control frequency and per-action shot count are varied under a realistic noisy backend, thereby identifying the regimes in which the quantum policy remains stable versus those in which it fails. As part of the same sweep, we also vary the training control frequency to probe robustness under train--deploy mismatch; however, its effect is secondary compared to the inference-time constraints. Overall, the resulting performance maps are intended as practical guidance for future real-time CartPole demonstrations on quantum hardware, where the achievable control frequency and measurement budget are jointly limited by end-to-end latency.

Finally, we address the dominant bottleneck for hardware deployment---the latency of standard cloud and software stacks---by implementing a low-level inference pipeline that bypasses the high-level interface and directly programs the control and readout electronics. Using this approach, we demonstrate inference iteration rates up to 6.2~Hz on a superconducting QPU, enabling substantially faster closed-loop policy execution than is achievable via standard access paths.

The remainder of this paper is organized as follows: section 2 details basic CartPole benchmarking task, the actor-critic framework, the hybrid network architecture, and the specific experimental protocols for simulation and hardware inference. Section 3 presents the benchmarking results and the latency analysis, followed by a discussion of the implications for real-time quantum control in Section 4.

\section{Methodology}

\subsection{The CartPole Environment}
\label{subsec:state_representation}
\begin{figure}
 \centering
        \includegraphics[width=0.45\textwidth]{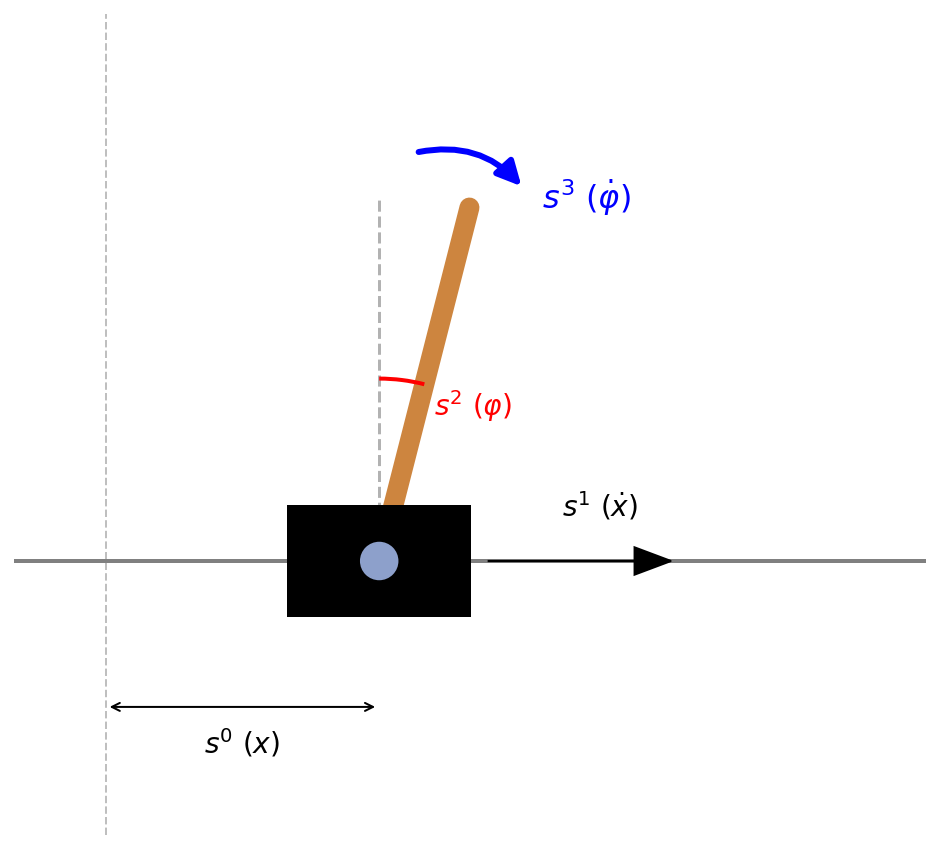}
 \caption{Schematic of the CartPole environment. The task consists of balancing the inverted pendulum mounted on the movable cart by applying discrete left or right forces. The observation states set $\tilde{s}=[s^1,s^2,s^3,s^4]$ is described by the cart position $x$ and velocity $\dot{x}$, as well as the pole angle $\varphi$ and angular velocity $\dot{\varphi}$ respectively.}

\label{fig:cart_pole_env}
\end{figure}

In the CartPole learning task, the objective is to stabilize an inverted pendulum mounted on a cart by applying discrete left or right forces at fixed control intervals, as illustrated in Fig.~\ref{fig:cart_pole_env}. An episode corresponds to a single interaction trajectory between the agent and the environment and terminates either when the pole exceeds a stability threshold or when a predefined maximum duration is reached. In this work, the maximum episode duration is set to correspond to 10 seconds in the environment.

Each episode is discretized into $N_{\text{t-steps}}$ control steps determined by the control frequency $f$. The agent receives a reward of one per timestep, such that the maximum achievable return is 500 at the default control frequency of $f = 50\,\mathrm{Hz}$, for example. At each timestep $i$, the environment provides a four-dimensional observation vector
\[
s_i = (x_i, \dot{x}_i, \varphi_i, \dot{\varphi}_i),
\]
encoding the cart position and velocity as well as the pole angle and angular velocity. The cart position and pole angle are bounded,
\begin{equation}
    x_i \in [-2.4,\,2.4], 
    \qquad 
    \varphi_i \in [-0.418,\,0.418]~\text{rad},
\end{equation}
while the velocity components $\dot{x}_i$ and $\dot{\varphi}_i$ are nominally unbounded. Aside from reaching the 10-second-mark,  episodes terminate if the cart leaves the interval $|x_i| > 2.4$ or the pole angle exceeds $|\varphi_i| > 0.418$ radians. The learning agent is rewarded +1 for each time step the pole stays within a smaller bound $|\varphi_i| > 0.2$ radians.

\iffalse
We measure sample efficiency by the number of training episodes required to satisfy the success criterion, defined as achieving the maximum episodic return of 500 for 100 consecutive episodes at $f=50$Hz.
\fi

\subsubsection{Feature Selection}
\label{sec:feature_selection}

In our experiments we do not feed the full state $s_i$ to the agent, but instead use a reduced three-dimensional feature vector
\begin{equation}
    \tilde{s}_i = \bigl(\dot{x}_i, \varphi_i, \dot{\varphi}_i\bigr),
    \label{eq:reduced_state}
\end{equation}
obtained by discarding the cart position $x_i$. This reduction is motivated by the simple single-qubit design of the hybrid agent (elaborated more upon in Sec.~\ref{subsec:hybrid_description}): the encoding of the VQC maps $\tilde{s}_i$ to a point on the Bloch sphere, which is effectively described by two real degrees of freedom (up to a global phase). Therefore, restricting the input state to three observables helps reduce the degeneracy of the state encoding. However, since the episodic termination criterion dictates a boundary on the cart position, the observable was kept track of and the episode was terminated whenever $x_i$ did not lie within the allowed region.

\iffalse
In a simple feature-selection study over subsets of $\{x_i,\dot{x}_i,\varphi_i,\dot{\varphi}_i\}$, the cart position $x_i$ was found to be the least informative for solving CartPole, and was therefore dropped in both the classical and hybrid models.
\fi

\subsection{The Actor-critic Method}
\label{subsec:cartpole_rl_bg}
%A reinforcement learning model is a decision-making process where the aim is to complete an objective. The model usually consists of an environment and a learning agent which acts on the environment. In our case, we refer to the CartPole benchmark as the environment. A reinforcement learning environment consists of a set of states $S$ and a set of actions $A$ which the agent can perform in order to change the environment state. At each  time-step $t_i$, with $i$ is the time-step number, the agent receives a new state $s_i \in S$ and picks an action $a_i \in A$. By performing an action and changing the environment state, the agent obtains a reward which evaluates the quality of that action based on the learning objective. The goal is to maximize the cumulative reward over the episode and the learned strategy to achieve this goal is called policy.

In the CartPole environment, for each time-step $t_i = i \Delta t$, where $\Delta t$ represents the duration of a single time step, the system state $s_i$ is defined by the \iffalse cart position $x$,\fi cart velocity $\dot{x}$, pole angle $\varphi$, and pole angular velocity $\dot{\varphi}$:
\begin{equation}
 s_i = \left( \dot{x}(t_i), \varphi(t_i), \dot{\varphi}(t_i) \right)
\end{equation}
The objective is to maximize the return. For a trajectory $\tau =(s_0,a_0,s_1,a_1,\dots,s_T,a_T)$ of length $T$, the return is defined as the cumulative discounted reward:

\begin{equation}
    R(\tau) = \sum_{i=0}^{T} \gamma^i r_i(s_i,a_i).
\end{equation}
where $\gamma$ is the discount rate. We implement the Actor-critic (AC) method with a policy gradient (PG) algorithm \cite{sutton1999policy}. AC contains an actor that takes the states as input and outputs the probability distribution of the action, while the critic evaluates the value of the chosen action. The actor learns the policy $\pi_{\theta}(a_i \mid s_i)$, which governs the probability of selecting action $a_i$ given state $s_i$ \cite{Sutton1998-tz}. To maximize performance, the policy parameter $\theta$ is updated via gradient descent from a surrogate loss function that approximates the negative expected return:
\begin{equation} \theta_{i+1} = \theta_i - \alpha \nabla L_{\text{actor}}(\theta_i) 
\end{equation}
Here, $\alpha$ is the learning rate and the loss function for the actor is defined as
\begin{equation}
    L_{\text{actor}}(s_i, \theta) = -\frac{1}{T} \sum_{i=0}^{T} \log(\pi_{\theta}(a_i \mid s_i)) \cdot A^{\pi_{\theta}}(s_i, a_i),
\end{equation}
where $A^{\pi_{\theta}}$ represents the advantage of the state-action pair at time-step $i$. In the Actor-Critic framework, the advantage is estimated using the Temporal Difference (TD) error:
\begin{equation} 
A_t \approx \delta_i = r_i + \gamma V_\phi(s_{i+1}) - V_\phi(s_i) 
\end{equation}
The state value function $V_\phi(s_i)$, parameterized by weights $\phi$ and computed by the critic network, estimates the expected cumulative discounted reward from state $s_i$:
\begin{equation}
V_\phi(s_i) = \mathbb{E}\left[ \sum_{k=0}^{\infty} \gamma^k r_{i+k} \mid s_i \right].
\end{equation}
Traditional reinforcement learning methods often rely on lookup tables for value estimation; however, this approach is computationally expensive for large state-action spaces. Neural networks are often employed here to approximate these values efficiently, accommodating non-linearities \cite{Israilov2023-ps}. The critic network is optimized to minimize the error between its prediction $V_\phi(s_i)$ and the TD target $y_i = r_i + \gamma V_\phi(s_{i+1})$. We employ the Huber loss function to effectively handle outliers:
\begin{equation}
L_{\text{critic}} = \frac{1}{T} \sum_{i=0}^{T} H_{\delta}(y_i - V_\phi(s_i))
\end{equation}

where

\begin{equation}
H_{\delta}(e) =
\begin{cases}0.5 e^2 & \text{if } |e| < \delta \\{\delta} (|e| - 0.5 \delta) & \text{otherwise}.
\end{cases}
\end{equation}

The reinforcement learning algorithm is summarized in Algorithm \ref{alg:ac_recursive}.

\begin{algorithm}[t]
\caption{Episodic Actor-Critic with Policy Gradient}
\label{alg:ac_recursive}
\begin{algorithmic}
\Require Actor $\pi_{\theta}$, Critic $V_{\phi}$, Optimizers $\text{Adam}_{\theta}, \text{Adam}_{\phi}$
\For{episode $e = 1 \dots E$}
    \State $t \leftarrow 0$, $\tau \leftarrow \emptyset$, Reset environment to $s_0$
    
    \While{not done and $t < T_{max}$} 
        \State Get Action and Value: $a_t \sim \pi_{\theta}(\cdot|s_t)$, $v_t \leftarrow V_{\phi}(s_t)$
        \State Execute $a_t$, observe $r_t, s_{t+1}, d_t$ \Comment{$d_t=1$ if terminal}
        \State Store $ \tau \leftarrow (r_t, v_t, \log \pi(a_t|s_t), d_t)$
        \State $t \leftarrow t+1$
    \EndWhile
    
    \State Initialize Returns list $D \leftarrow \emptyset$ 
    \State $R_{next} \leftarrow V_{\phi}(s_t)$
    \For{$k = t-1$ \textbf{down to} $0$}
        \State $R_k \leftarrow r_k + \gamma \cdot R_{next} \cdot (1 - d_k)$
        \State Insert $R_k$ at start of $D$
        \State $R_{next} \leftarrow R_k$
    \EndFor
    
    \State \textbf{Compute Advantage:} $A_k = R_k - v_k$ \Comment{Using stored $v_k$}
    
    \State \textbf{Update Networks:}
    \State $L_{\text{actor}} = - \frac{1}{T} \sum_{k=0}^{T} A_k \log \pi_{\theta}(a_k|s_k)$
    \State $L_{\text{critic}} = \frac{1}{T} \sum_{k=0}^{T} H_{\delta}(R_k - v_k)$
    
    \State $\theta \leftarrow \text{Adam}_{\theta}(\nabla_{\theta} L_{\text{actor}})$; $\phi \leftarrow \text{Adam}_{\phi}(\nabla_{\phi} L_{\text{critic}})$
\EndFor
\end{algorithmic}
\end{algorithm}

\FloatBarrier

\subsection{Classical Neural Network Model}
\label{subsec:cnn_description}
As a classical baseline, we represent both the actor and the critic by separate fully connected neural networks. Each network takes the continuous state vector $s_i$ as input and processes it through two hidden layers with 128 and 256 units, respectively. Both hidden layers use rectified linear unit (ReLU) activations to capture nonlinear dependencies in the state space. The actor and critic networks are shown in Figure 

\begin{figure}[t]
 \centering
        \includegraphics[width=0.8\textwidth]{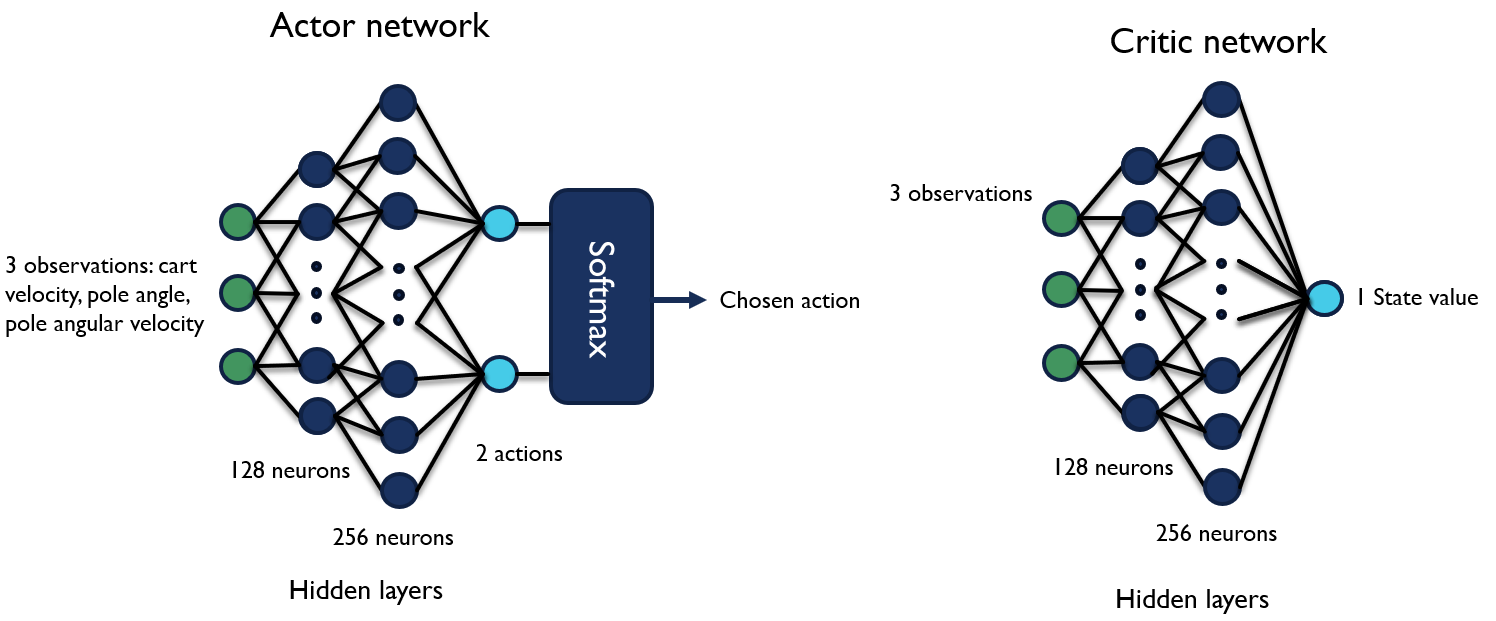}
 \caption{Classical actor network (left) and critic network (right) illustrations. Both networks share an identical internal structure consisting of an input layer for observations followed by two hidden layers of 128 and 256 neurons, respectively. The input layer received the raw observations. While the actor network contained the Softmax function to calculate the action probability distribution, the critic outputs a single state value judging how well the current environment states is. By comparing this value to actual reward, the agent determines whether the actor's chosen action was better or worse than expected. }
\label{fig:CNN_actor_critic}
\end{figure}

The actor network maps the input state to a vector of logits in $\mathbb{R}^{|A|}$, where $|A|=2$ is the number of discrete actions. These logits are passed through a softmax layer to obtain a categorical policy $\pi_{\theta}(a_i \mid s_i)$ over actions. In contrast, the critic network has an identical hidden-layer structure but terminates in a single linear output neuron, producing a scalar estimate of the state value $V_{\phi}(s_i)$. The actor and critic do not share parameters, allowing each to specialize in policy representation and value estimation, respectively.\ref{fig:CNN_actor_critic}.

\subsection{Quantum-classical Hybrid Neural Network}
\label{subsec:hybrid_description}

\begin{figure}[t]
 \centering
        \includegraphics[width=0.7\textwidth]{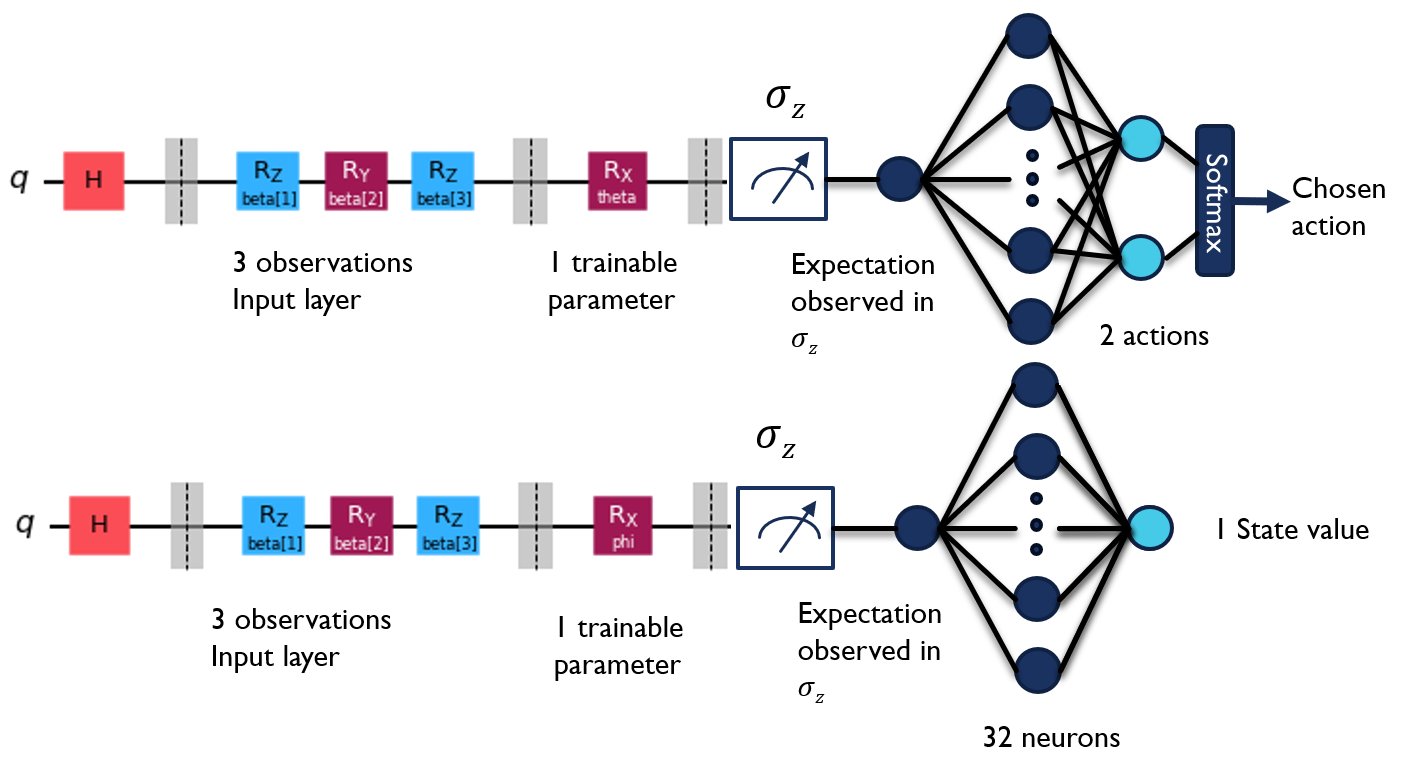}
 \caption{Hybrid quantum neural network for the actor (top) and critic (bottom). The functions of both actor and critic remain the same as its classical counterpart. The NN has been replaced by PQCs to approximate policy function (actor) and value function (critic). Each PQC output is copied 32 times before fully connected to two outputs (actor) and one output (critic). Soft is also used to output the final probability distribution.}
\label{fig:QNN_actor_critic}
\end{figure}

The hybrid quantum--classical network architecture employs a single-qubit parameterized quantum circuit connected to classical, fully connected layers adapted from~\cite{Hsiao2022-uj}. The quantum module begins with a Hadamard gate to create the $|+\rangle$ state, followed by a three-rotation state-encoding sequence ($R_z$--$R_y$--$R_z$) that maps the reduced CartPole observation vector (given in Sec.~\ref{eq:reduced_state}) onto the Bloch sphere, and concludes with a trainable $R_x(\theta)$ rotation. 
The resulting expectation value of a Pauli-$Z$ measurement is copied to each input neuron of the downstream classical actor and critic networks, each consisting of a single hidden layer with 32 neurons and ReLU activation. The entirety of the hybrid quantum neural network reinforcement learning model for the CartPole task is illustrated in Figure \ref{fig:QNN_actor_critic}.

We implement a variational quantum circuit (VQC) to leverage this high-dimensional space for encoding reinforcement learning states \cite{Cerezo2021-yd}. At each time step $i$, the reduced observable vector described in Sec. \ref{subsec:state_representation} is first preprocessed element-wise with the $\arctan$ function to further reduce degeneracy along each axis individually. This is indicated as $\beta = \arctan(x_i,\dot{x}_i,\varphi_i,\dot{\varphi}_i) $ in Figure \ref{fig:QNN_actor_critic}. The preprocessed states are then encoded into a quantum state $|\psi(s_i)\rangle$ via a unitary embedding operator $U(s_i)$:

\begin{equation}
    |\psi (s_i)\rangle = U(s_i) |0\rangle
\end{equation}
where $|0\rangle$ represents the initial quantum state. For this specific task, we utilize a single-qubit architecture. The embedding operator $U(s_i)$ comprises a Hadamard gate to induce superposition, followed by rotation gates ($R_z, R_y, R_z$).
\begin{equation}
    U(s_i) = R_z(\beta_3) R_y(\beta_2) R_z(\beta_1) H.
\end{equation}
Following encoding, the circuit applies a rotation gate around the X-axis $R_x(\theta)$, where $\theta \in \mathbb{R}$ represents the trainable parameter whose initial value is controlled by the run-specific seed. The resulting quantum state $|\psi (s_i,\theta)\rangle$ is measured via projection onto the Pauli-Z axis ($\sigma_z$). The output of the VQC is given as follows:
\begin{equation}
    f(s_i,\theta) = \langle 0|U^\dagger(s_i,\theta)M_{\sigma_z} U(s_i,\theta)|0\rangle.
\end{equation}

The above expectation value can be simulated analytically or obtained as a statistical estimate by sampling the circuit $N_{\text{shots}}$ times. As the system size grows, concerns of exponential concentration \cite{larocca2025barren} rise for the latter approach, as exponentially many samples may be needed for the accurate estimation of the observable motivating a careful design of the circuit used.

The output $f(s_i, \theta)$ is repeated 32 times and passed to a classical feedforward neural network (FNN), that outputs probabilities for each possible action. The FNN consists of a layer of 32 neurons outputting 2 logits, which is followed by a softmax activation function. The process of reusing PQC output has been reported by ~\cite{Hsiao2022-uj} to increase convergence speed of the learning agent on CartPole benchmark. The critic outputs, as in the classical baseline, a single value. This architecture forms the basis for the training-inference compatibility study detailed in Section \ref{subsubsec: training-inference}, where we evaluate the model's robustness to the shot-noise and latency constraints inherent to the physical QPU. Finally, to enable gradient-based optimization of the quantum circuit parameters within the classical training loop, we utilize the parameter-shift rule. A detailed derivation of this method and its integration with standard backpropagation is provided in Appendix \ref{subsec:gradients}.

\FloatBarrier

\subsection{Experiment Categories and Backends}
\label{subsec:training}

In this work we consider three categories of experiments: (i) a noiseless benchmark comparing classical and hybrid quantum–classical agents, (ii) a training–inference compatibility study for the hybrid agent across different execution backends, control frequencies, and shot counts, and (iii) low–latency inference on a real quantum processor. All experiments use the CartPole-v1 environment with a maximum episode length corresponding $10$ seconds.

\subsubsection{Noiseless Baseline: Classical vs.\ Hybrid Agents}

The first set of experiments establishes a baseline comparison between the purely classical actor–critic agent with policy gradient described in Section~\ref{subsec:cnn_description} and two variants of the hybrid quantum–classical agent explained in Section~\ref{subsec:hybrid_description} by utilizing the default control frequency of $f_{\text{c}}=50$Hz. All agents are trained using the Qiskit BasicSimulator backend with 1024 shots.

For each model type (classical, hybrid with parameter–shift gradients, and hybrid with analytical gradients), we train an ensemble of $50$ independent agents. Agents differ only by their random initialization: classical and quantum network parameters are drawn from separate random seeds, and episode-wise environment initial states are also randomized. This ensemble setting allows us to probe the robustness of convergence behaviour rather than a single representative run.

Episodes are capped at $500$ time steps and training runs for at most $1500$ episodes. As a success criterion we require that an agent achieves an average episodic return of $500$ over $100$ consecutive episodes, which corresponds to near–perfect balancing performance. Once this criterion is met, training for that agent is terminated and the corresponding network parameters are stored for subsequent analysis. In this noiseless setting the hybrid agents exhibit faster convergence to this success criterion than the classical baseline.

\subsubsection{Training--Inference Compatibility Study for the Hybrid Agent}
\label{subsubsec: training-inference}
The second group of experiments investigates the transferability of learned hybrid quantum--classical policies across different training and deployment condition combinations. This compatibility study systematically evaluates how agents trained at a given control frequency perform when deployed at different inference frequencies and shot budgets.

All training experiments in this study are performed using the Qiskit \texttt{BasicSimulator} backend, which provides finite-shot sampling without additional device noise. Expectation values required for gradient estimation are obtained using 4096 measurement shots per circuit execution, enabling stable parameter-shift gradients while remaining computationally feasible. Agents are trained at five discrete control frequencies,
\[
f \in \{20,\,25,\,33,\,50,\,100\}\,\text{Hz},
\]
corresponding to time discretizations $\tau = 1/f \in \{0.05,\,0.04,\,0.03,\,0.02,\,0.01\}$ seconds. To ensure that all agents solve an equivalent control task, the episode duration is fixed to 10 seconds of simulated time across all frequencies by setting the maximum number of time steps according to $T_{\max} = 10/\tau$. As a result, agents experience different numbers of decision points (e.g.\ 200 steps at 20~Hz versus 1000 steps at 100~Hz) while facing the same physical control objective.

For each training frequency, an ensemble of ten independent agents is trained using unique, deterministic random seeds to ensure reproducibility. Training proceeds for 500 episodes using the actor--critic algorithm with policy gradients. Both quantum circuit parameters and classical network weights are optimized using the Adam optimizer with a learning rate of $\alpha = 0.05$ for both actor and critic ($\beta_1 = 0.9$, $\beta_2 = 0.999$). The hybrid architecture consists of a single-qubit variational circuit coupled to a classical neural network with 32 hidden units, as detailed in Section~\ref{subsec:hybrid_description}. Upon completion of training, the model weights including both quantum parameters and classical weights are saved for the subsequent compatibility evaluation.

The compatibility analysis evaluates each of the five training configurations across all inference frequencies, constructing a complete $5 \times 5$ transfer matrix for each inference shot count. During inference, the policies are executed using shot counts of 128, 256, 512, and 1024 shots per circuit execution, reflecting the execution time constraints typical of near-term quantum hardware. No further parameter updates are performed during inference and episode durations are again fixed to 10 seconds of simulated time by adjusting the timestep limit according to the inference frequency, enabling direct comparison of control performance across configurations. For each combination of training frequency, inference frequency, and shot count, 10 evaluation episodes are performed and the performance is quantified using mean duration. 

This experiment subset enables a quantitative assessment of several deployment-relevant questions for PQC-based reinforcement learning: the extent to which policies potentially overfit to a specific control frequency, the impact of frequency mismatch on closed-loop stability, and, most importantly the sensitivity of policy performance to both reduced sampling precision inference control frequency.

\subsubsection{Quantum Hardware Implementation for Low-Latency Inference}
The third set of experiments focuses on low–latency control on a real quantum processing unit. Here we no longer update the policy parameters; instead, we deploy a previously trained hybrid agent in pure inference mode and measure its control performance and iteration speed on hardware.

For these experiments we select an agent which has satisfied the stability criterion during training and demonstrated robust performance in the compatibility study when evaluated on the \texttt{IQMFakeAdonis} backend, which emulates the target device.

The selected agent is then executed on VTT Q5 using the optimized low–level control stack described in Section~\ref{subsec:hw_setup}. Episode scores and timing statistics (iteration rate and throughput) are recorded for different shot counts, providing a direct measurement of how the learned policies perform and how close current hardware can come to the real–time control requirements of the CartPole task.

\subsection{Training Details}

For the training-inference compatibility studies, the training configurations explored a comprehensive parameter space:
\begin{itemize}
    \item Gradient estimation: parameter-shift rule \cite{schuld2019evaluating} for shot-based backends; analytical gradients via automatic differentiation for ExpVal backend
    \item Measurement shots per circuit execution: $N_{\text{shots}} \in \{128, 256, 512, 1024\}$; analytical expectation values for ExpVal backend
    \item Control frequencies: $f \in \{20, 25, 33, 50, 100\}$ Hz, corresponding to time discretization parameters $\Delta t \in \{0.05, 0.04, 0.03, 0.02, 0.01\}$ seconds
\end{itemize}

For the optimization procedure employed the Adam optimizer \cite{Kingma2014-or} with separate learning rates for actor and critic networks. Quantum circuit parameters $\theta$ and episode-wise environments were initialized using agent-specific random seeds, yielding distinct circuit and environment initializations for each agent and thereby promoting diverse training trajectories across the ensemble.

\iffalse
Model stability was defined as the achievement of ten consecutive episodes with maximum reward (500 timesteps in the CartPole environment). This criterion ensured that saved configurations represented converged solutions rather than transient high-performance states. Upon reaching stability, the complete model state was preserved, including the quantum circuit parameters $\theta_{\text{actor}}$ and $\theta_{\text{critic}}$ as well the classical network weights.

\fi

\subsection{Low-latency Inference Configuration}
\label{subsec:hw_setup}
We run our experiments on VTT Q5, a 5-qubit superconducting quantum computer developed by IQM and VTT. The qubits are flux-tunable and assume a star topology. The native gates consists of the single-qubit Phased-$RX$ (PRX) gate, the $CZ$-gate, and single-qubit measurements in the $Z$-basis. The PRX gate is defined as 
\begin{equation}
PRX(\phi, \theta) = e^{-i(X \cos \phi + Y \sin \phi) \theta/2}
\end{equation}
and enables arbitrary rotations around the $X$/$Y$-plane. The gate is implemented as a microwave pulse. Rotations around the $Z$-axis are achieved by virtual-$Z$ rotations \cite{McKay-2016}, corresponding to phase shifts in the control electronics. The PRX gate is implemented as a Gaussian DRAG pulse~\cite{Motzoi-2009} with a duration of 120~ns. The circuit in Figure~\ref{fig:QNN_actor_critic} can be expressed by three PRX pulses

\begin{equation}
    PRX\left(\theta, -\beta_1 - \beta_3\right)\;\;PRX\left(\beta_2, \frac{\pi}{2} - \beta_1\right)\;\;PRX\left(\frac{\pi}{2}, \frac{\pi}{2}\right).
\end{equation}

\begin{figure}[t]
  \centering
  \includegraphics[width=0.8\textwidth]{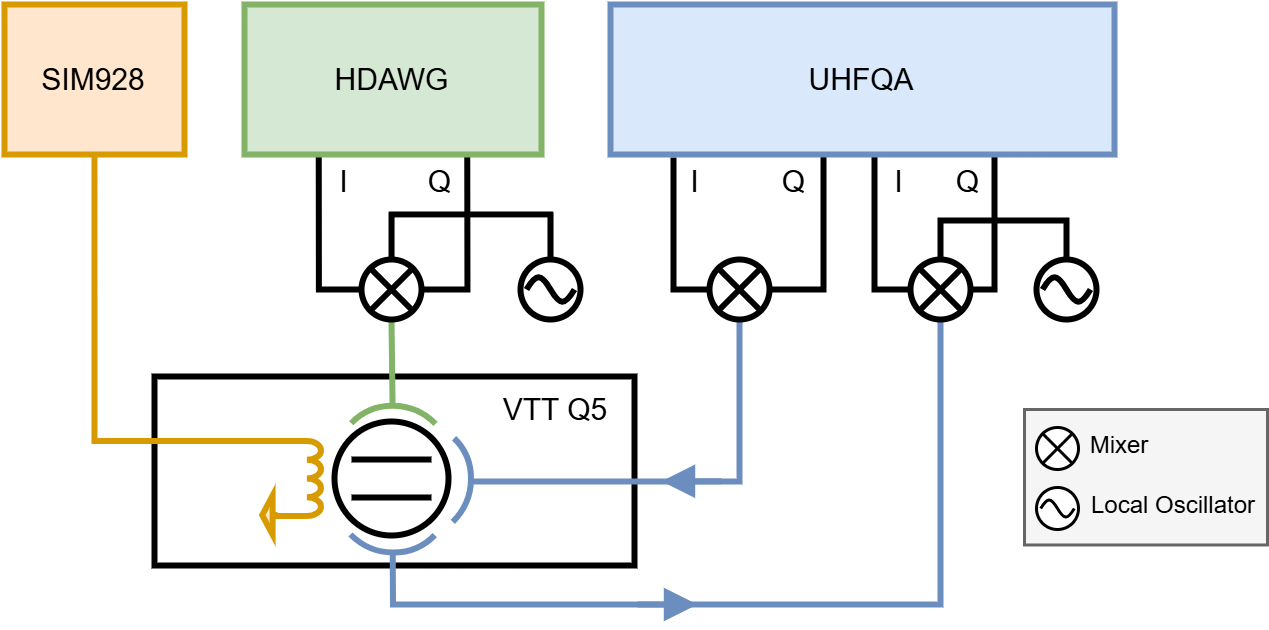}
  \caption{Experimental setup for measurements on Qubit~3 of the VTT~Q5 device. The qubit flux bias is provided by an SRS SIM928 source (orange), which sets the qubit frequency and remains constant throughout the experiment. Microwave drive pulses are generated by the HDAWG (green), while readout pulses are produced by the UHFQA. Qubit drive pulses are implemented via frequency up-conversion, where narrow-band intermediate-frequency in-phase (I) and quadrature (Q) signals are mixed with a high-frequency local oscillator using an IQ mixer.}
  \label{fig:q5_setup}
\end{figure}

The IQM quantum control stack (QCS) accepts Qiskit circuits as well as pulse sequences, defined in IQM's pulse-level language, and compiles them to hardware instructions using the instrument drivers. For VTT Q5, an arbitrary waveform generator (AWG) from Zurich Instruments (HDAWG) is used to drive the qubits, which generates pulses at a sampling rate of 2.4~GSa/s. For readout, an ultra-high frequency quantum analyzer (UHFQA) is used, with a sampling rate of 1.8~GSa/s. Figure~\ref{fig:q5_setup} shows the experimental setup. Both HDAWG and UHFQA are programmed in Zurich Instruments' SeqC language, allowing precise playback of a sequence of microwave pulses on all channels. 

To execute our parameterized circuit in Qiskit, we first bind the free gate parameters $(s^0_i, s^1_i, s^2_i, \theta_i)$ and then transpile the circuit to native gates. The circuit is then submitted to the QCS and transformed into an intermediate representation (IR). The IR is translated into a sequence of pulses on each channel (e.g., qubit drive and flux bias). The parameters of each gate are translated to pulse parameters (e.g., amplitude and duration) determined by daily calibration (see Table~\ref{tab:hardware-details} for reference values). In a final step, the waveform samples for each pulse and the SeqC code for the HDAWG and UHFQA are generated and uploaded to the devices. Finally, execution is triggered. After successful execution, the results are fetched from the UHFQA by the QCS and returned in Qiskit format to the user.

In our experiments utilizing the standard Qiskit software stack we found that the execution speed is limited by the re-compilation of SeqC code and upload of code and waveforms to the devices upon circuit parameter changes. An average step frequency of 0.14 Hz for 512 shots (Table~\ref{tab:iteration-speeds}) was achieved. To speed up the execution, we bypass the QCS and connect directly to the Zurich Instrument devices. The execution on the Zurich Instrument devices is driven by a command table (CT). The CT consists of an indexed list, where each entry points to a waveform stored in memory and additional phase-shift and amplitude-scaling values for the IQ-components of the pulses. The SeqC code consists of triggers executing CT entries and idling instructions controlling the timing of the pulses for driving and measuring the qubit.
For our parameterized circuit this timing is fixed. Therefore, the SeqC code does not require re-compilation and we can skip this expensive step. For each new set of parameters we therefore only upload new waveforms and an updated command table and trigger the execution. This simple but effective solution reduces the execution time significantly as shown in Table~\ref{tab:iteration-speeds}. 
To further speed up the execution we optimized the wait time between two shots. To ensure that the qubit is in the ground state at the beginning of the execution, a wait time $\gg T_1$ is prepended to each circuit. The default on VTT Q5 is $398$ $\mu$s. Sweeping over the reset wait time we found $220$ $\mu$s to be optimal, shortening the execution time significantly while maintaining circuit fidelity (see Figure~\ref{fig:sweep_reset_wait_time}) .

\FloatBarrier

\section{Results}
\label{sec:results}

Overall, our experiments show that (i) the single-qubit hybrid agent can solve CartPole with substantially fewer episodes than a classical actor--critic baseline under ideal simulation even with the use of parameter-shift, (ii) the average balancing duration during inference increases with $f_{\text{c,inf}}$ but not necessarily with $f_{\text{c,train}}$, and (iii) low-level access to the quantum control stack yields over an order of magnitude reduction in iteration latency on real hardware while achieving successful inference test scores.

\subsection{Learning Convergence}
\label{sec:learning_convergence}

\begin{figure}[htbp]
  \centering
  \includegraphics[width=1.02\textwidth]{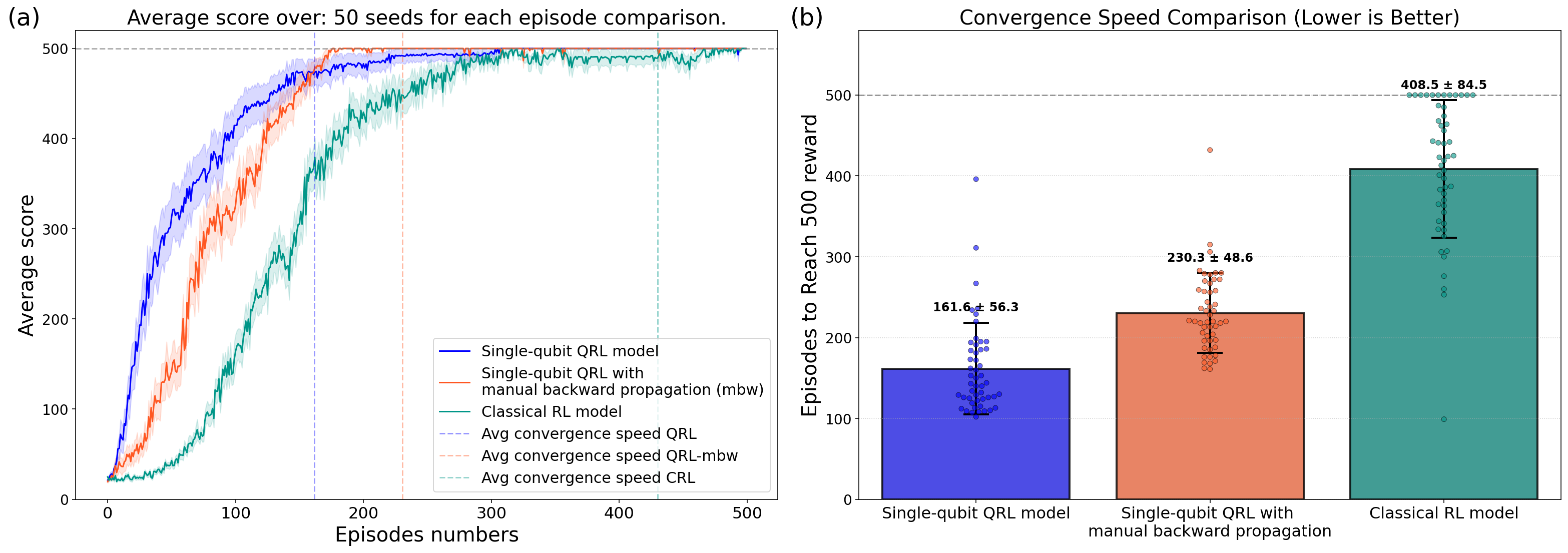}
  \caption{Average reward scores over 50 runs (5a). The single-qubit QRL model (blue) achieves the maximum reward threshold significantly faster than the Classical RL baseline (green), validating the higher sample efficiency observed in simulation. The orange curve represented the same QRL model with manual backward propagation process implemented. The dashed lines show the average episodes taken to solve the CartPole task. Average episode counts needed for the single-qubit QRL and classical RL models to reach the success criterion of 100 consecutive perfect episodes (5b). The Single-Qubit model (blue) demonstrates a ~2.7x improvement in sample efficiency (161.6 ± 56.3 episodes) compared to the Classical model (429.4 ± 117.9 episodes) under ideal simulation conditions. Each dot represent the episode number when the models solve the task at different seed.}
  \label{fig:score_comparison}
\end{figure}

\begin{figure}[h]
\centering
\begin{minipage}{0.45\textwidth}
\centering
\includegraphics[width=\textwidth]{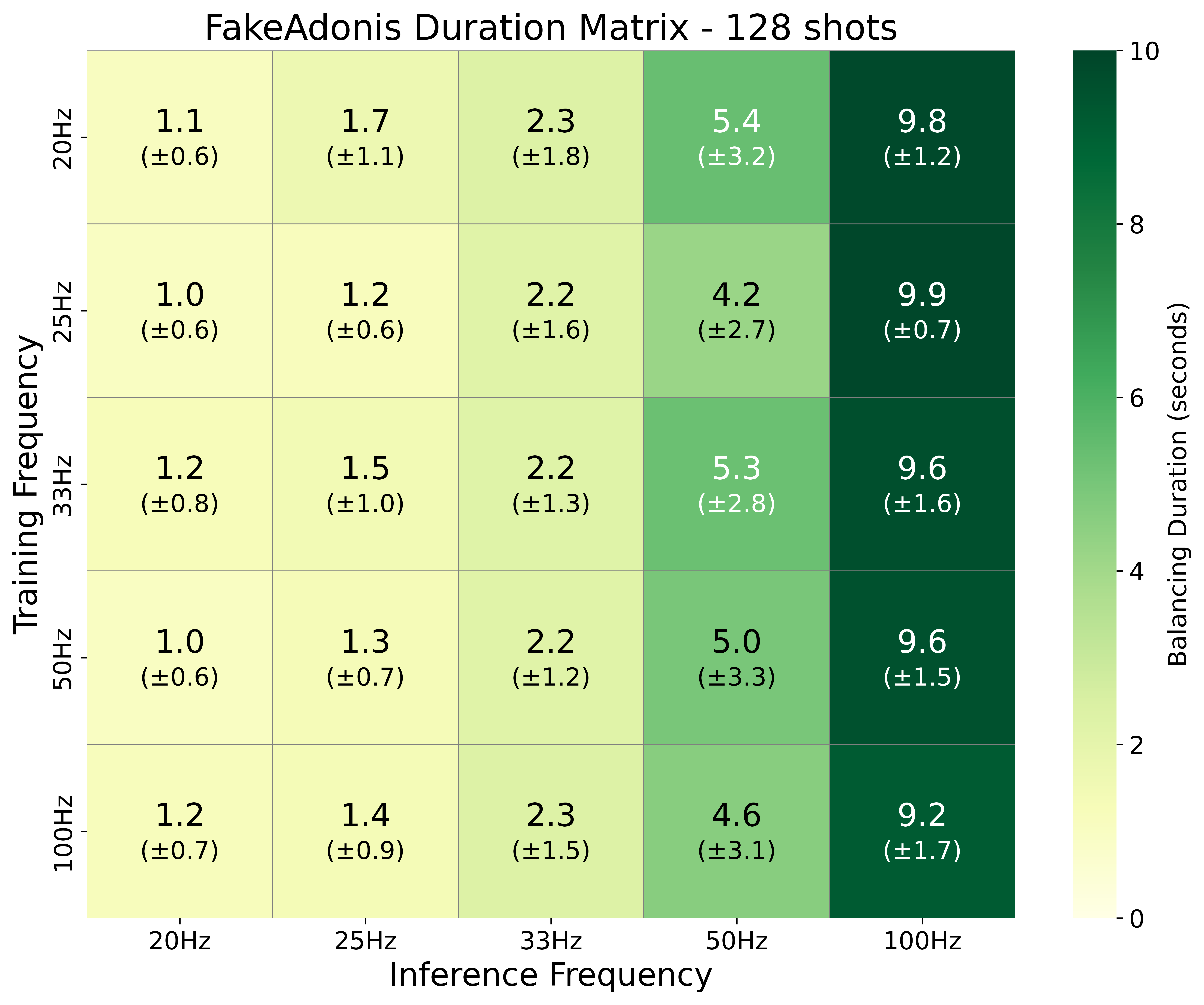}\\
\small (a)
\end{minipage}
\hfill
\begin{minipage}{0.45\textwidth}
\centering
\includegraphics[width=\textwidth]{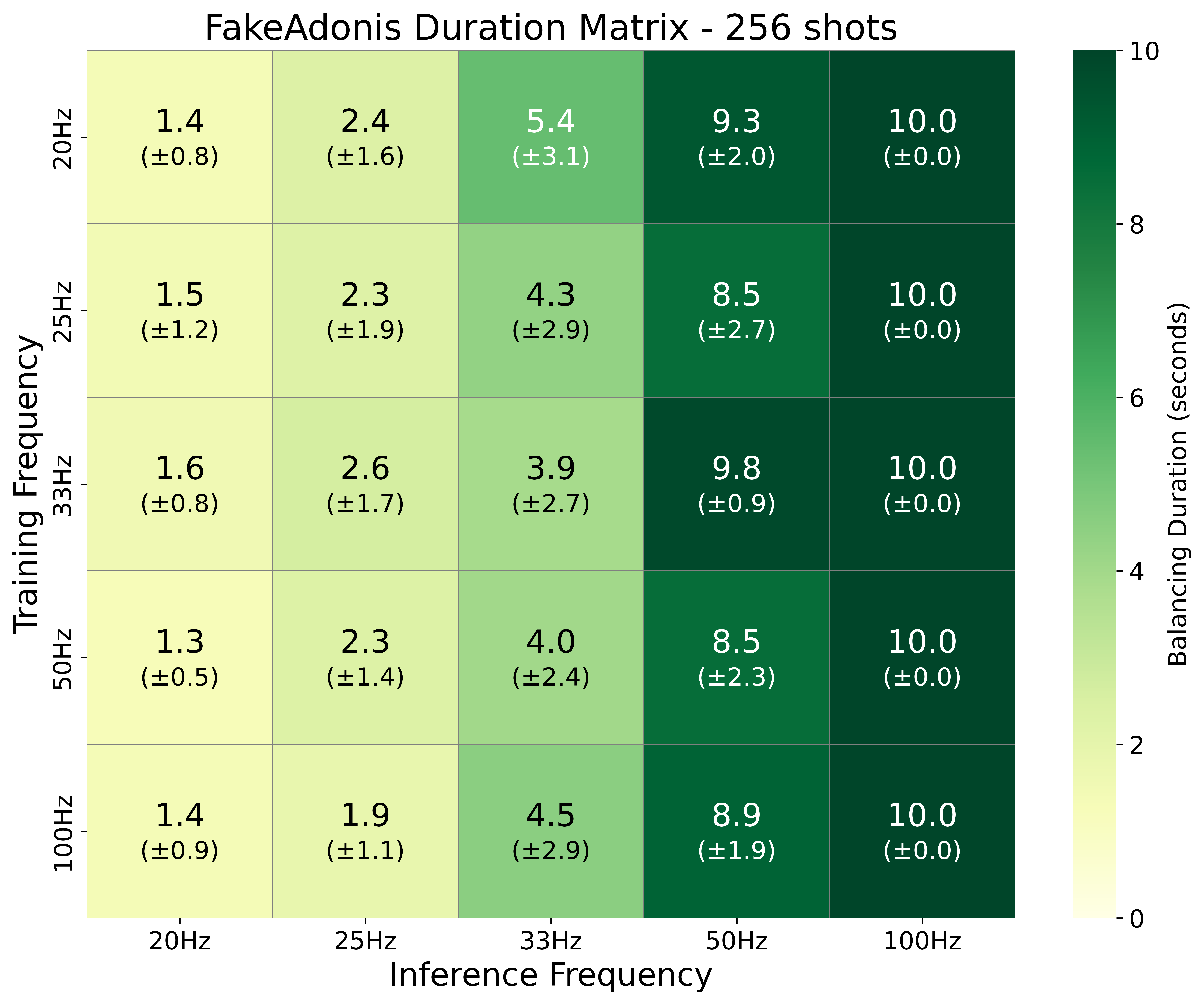}\\
\small (b)
\end{minipage}
\caption{FakeAdonis duration matrices comparing 128 shots (\textbf{a}) and 256 shots (\textbf{b}). The heatmaps show mean balancing duration in seconds ($\pm$ standard deviation) across different training and inference frequency combinations.}
\label{fig:fakeadonis_128_256shots}
\end{figure}

\begin{figure}[t]
\centering
\begin{minipage}{0.45\textwidth}
\centering
\includegraphics[width=\textwidth]{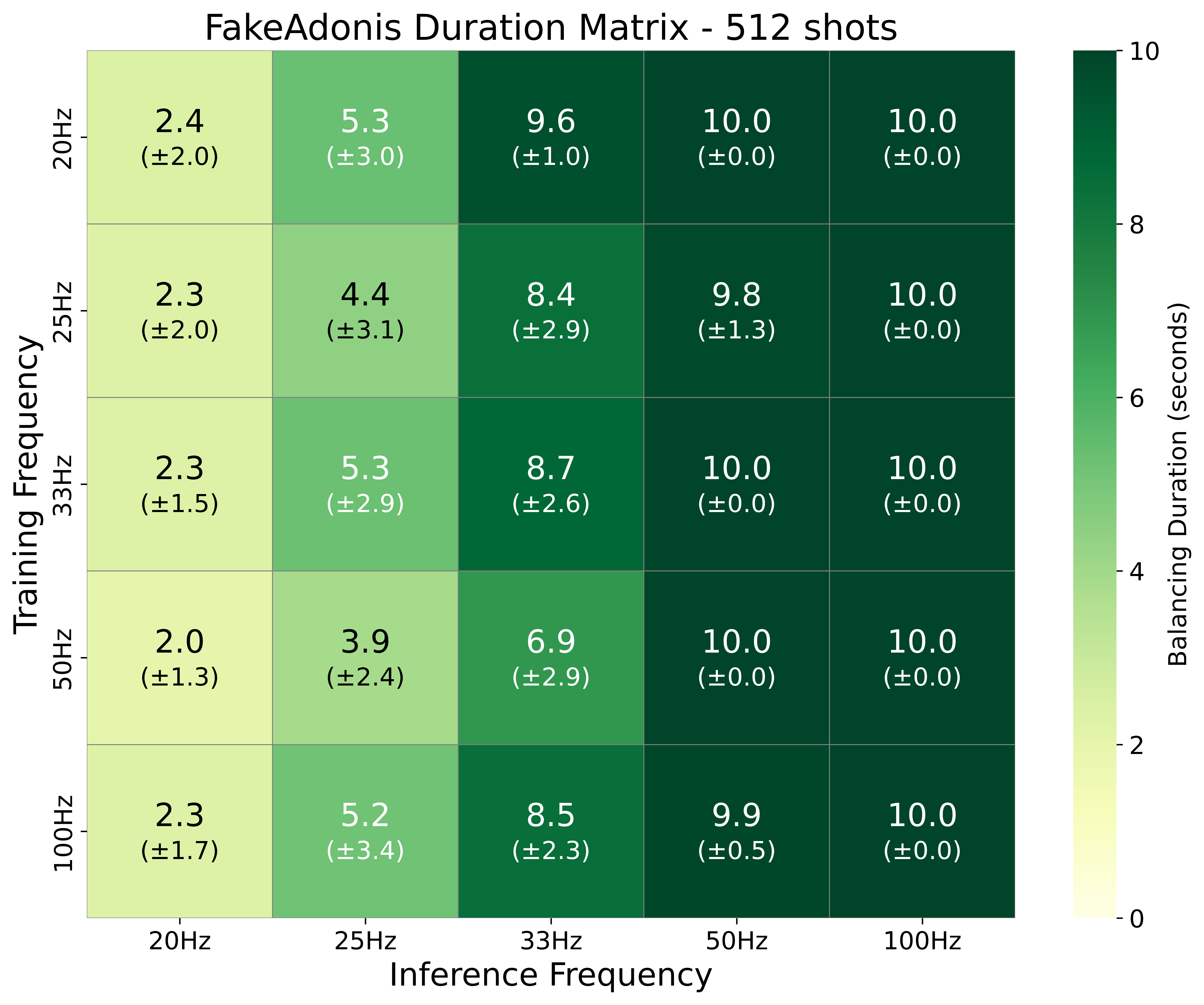}\\
\small (a)
\end{minipage}
\hfill
\begin{minipage}{0.45\textwidth}
\centering
\includegraphics[width=\textwidth]{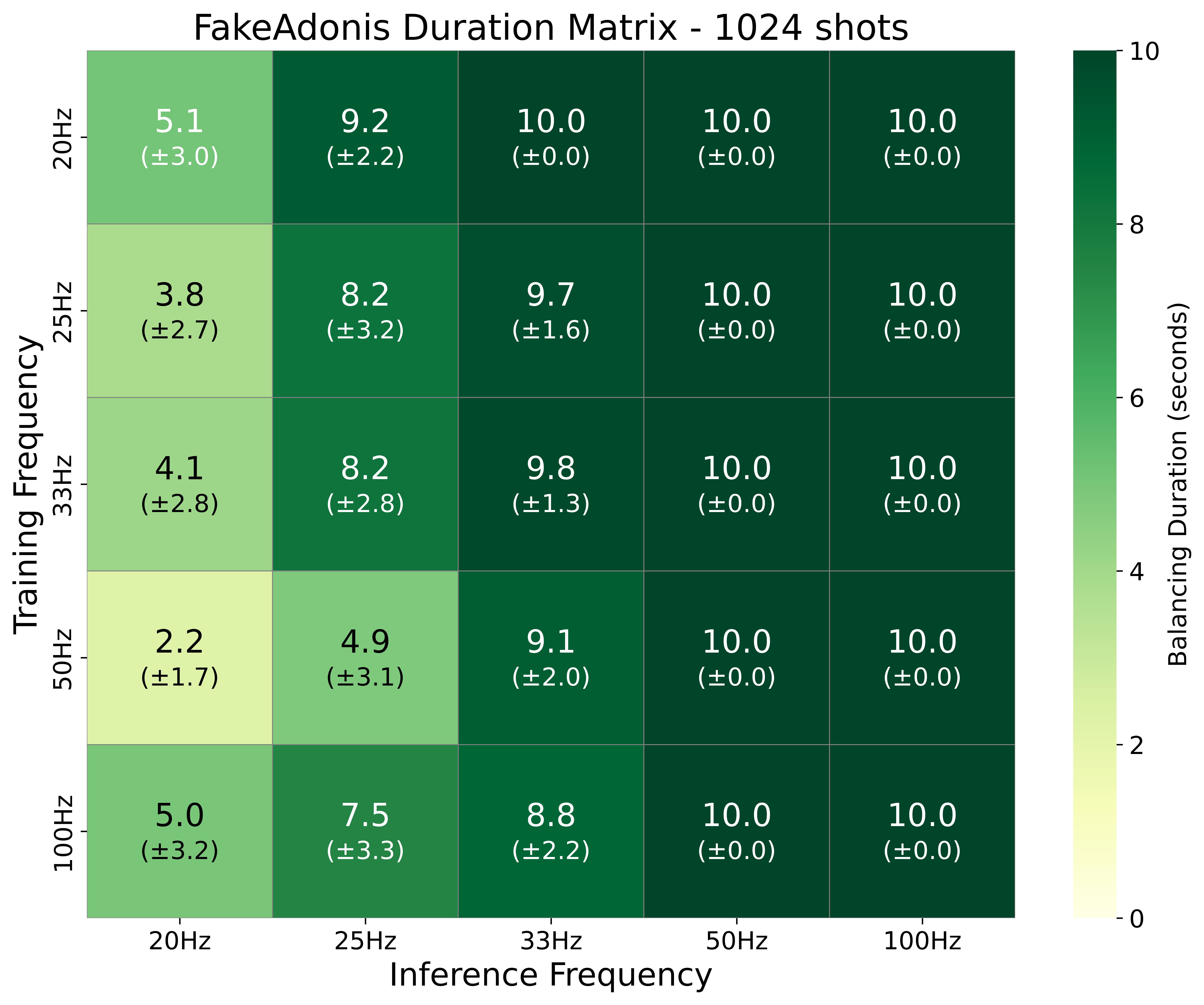}\\
\small (b)
\end{minipage}
\caption{FakeAdonis duration matrices comparing 512 shots (\textbf{a}) and 1024 shots (\textbf{b}). Higher shot counts provide more stable quantum measurements, resulting in improved balancing performance and reduced variance across training--inference frequency combinations.}
\label{fig:fakeadonis_512_1024shots}
\end{figure}

We first compare learning dynamics of the classical and hybrid agents in expectation-value simulation (Figures~\ref{fig:score_comparison}). 
We then analyse training--inference compatibility across backends, shot counts, and control frequencies using a series of balancing duration matrices (Figures~\ref{fig:fakeadonis_128_256shots} and \ref{fig:fakeadonis_512_1024shots}). 
Finally, we present real-hardware inference benchmarks on VTT's IQM five-qubit superconducting QPU, including iteration rates and corresponding episode scores (Table~\ref{tab:iteration-speeds}, Figure~\ref{fig:iteration_performance_scores}).

Figure~\ref{fig:score_comparison} shows the episode-wise return trajectories and the average episode counts required to solve the task for the hybrid agent (with both analytical and parameter-shift gradient estimation) and for the fully classical agent. Results are averaged over 50 independently seeded runs of 500 episodes at the CartPole-v1 default control frequency of $f = 50~\text{Hz}$. The hybrid agent trained with analytical gradients solves the task in the fewest episodes and maintains the most stable performance thereafter. Both hybrid variants converge significantly faster than the classical baseline, as shown in Fig.~\ref{fig:score_comparison}b.

\iffalse
\begin{figure}[htbp]
  \centering
  \includegraphics[width=0.5\textwidth]{Figures/convergence_speed_comparison_bar_chart.png}
  \caption{Average episode counts needed for the Single-Qubit QRL and Classical RL models to reach the success criterion of 100 consecutive perfect episodes. The analytical Single-Qubit model (blue) demonstrates a ~2.7x improvement in sample efficiency (161.6 ± 56.3 episodes) compared to the Classical model (429.4 ± 117.9 episodes) under ideal simulation conditions. Each dot represent the episode number when the models solve the task at different seed.}
  \label{fig:bar_chart}
\end{figure}
\fi

\subsection{Effects of Control Frequency and Shot Count for Real-Time Inference}

The second experiment category quantifies how hybrid policies behave under deployment constraints that may differ from training. This serves two roles in this work. First, it guides the choice of training configuration and inference settings used for the real-QPU experiments in Section~\ref{sec:hw_results}. Second, it provides concrete parameter-regime guidance for future real-time demonstration consideration, where the achievable control frequency and inference shot budget are jointly limited by latency. Increasing the control frequency simultaneously increases the number of decision points per episode, while increasing the shot count improves estimation accuracy at the cost of execution latency. As a result, policy performance at inference time is governed by a trade-off between temporal resolution and measurement precision.

\iffalse
A key practical question for deploying RL controllers on quantum hardware is how the learned policy behaves as control frequency and sampling resources are varied at inference time. In principle, the control frequency available on a given QPU can be anticipated in advance and matched during training. However, increasing the control frequency simultaneously increases the number of decision points per episode and amplifies the impact of finite-shot noise on the inferred policy outputs, while increasing the shot count improves estimation accuracy at the cost of execution latency. As a result, policy performance at inference time is governed by a trade-off between temporal resolution and measurement precision.
\fi

Motivated by this trade-off, we systematically explore how policies trained at different control frequencies perform when evaluated across a range of inference-time frequencies and shot budgets. All agents are trained using the Qiskit \texttt{BasicSimulator} backend under pure statevector with 4096 shots, while inference is carried out with varying control frequencies and shot counts on the \texttt{FakeAdonis} backend with a noise model that emulates the target QPU of the VTT Q5. This empirical study is designed to identify training–inference combinations that yield reliable control performance under realistic execution constraints, and to assess whether policies trained at higher control frequencies—which offer finer temporal resolution during training—exhibit increased sensitivity to finite-shot and device-level noise during deployment.

\iffalse
\begin{figure}[t]
\centering
\begin{minipage}{0.45\textwidth}
\centering
\includegraphics[width=\textwidth]{Figures/fakeadonis_duration_matrix_2048shots.png}
\end{minipage}
\hfill
\begin{minipage}{0.45\textwidth}
\centering
\includegraphics[width=\textwidth]{Figures/fakeadonis_duration_matrix_4096shots.png}
\end{minipage}
\caption{FakeAdonis duration matrices comparing 2048 shots (left) and 4096 shots (right). The progression shows improved measurement precision and reduced quantum noise with increasing shot counts at the highest precision levels.}
\label{fig:fakeadonis_2048_4096shots}
\end{figure}
\fi

\FloatBarrier

\subsection{Low-latency Inference on a Real QPU}
\label{sec:hw_results}

We next benchmark inference-only execution of a trained hybrid policy on VTT Q5 and quantify the latency benefit of bypassing the standard software stack. We compare the default IQM execution path against the low-level inference pipeline explained in Sec.~\ref{subsec:hw_setup} that directly drives the control and readout electronics.

Table~\ref{tab:iteration-speeds} reports the average iteration rates (policy evaluations per second) for shot counts $N_{\text{shots}}\in\{128,256,512,1024\}$. The standard IQM stack achieves $\approx 0.14$~iter/s across regardless of shot count, whereas the low-level path reaches 6.23, 5.62, 4.28, and 2.71~iter/s, corresponding to speedups of $43.3\times$, $39.3\times$, $30.1\times$, and $18.8\times$, respectively.

\begin{figure}[!h]
  \centering
  \begin{minipage}{0.48\textwidth}
    \centering
    \includegraphics[width=\textwidth]{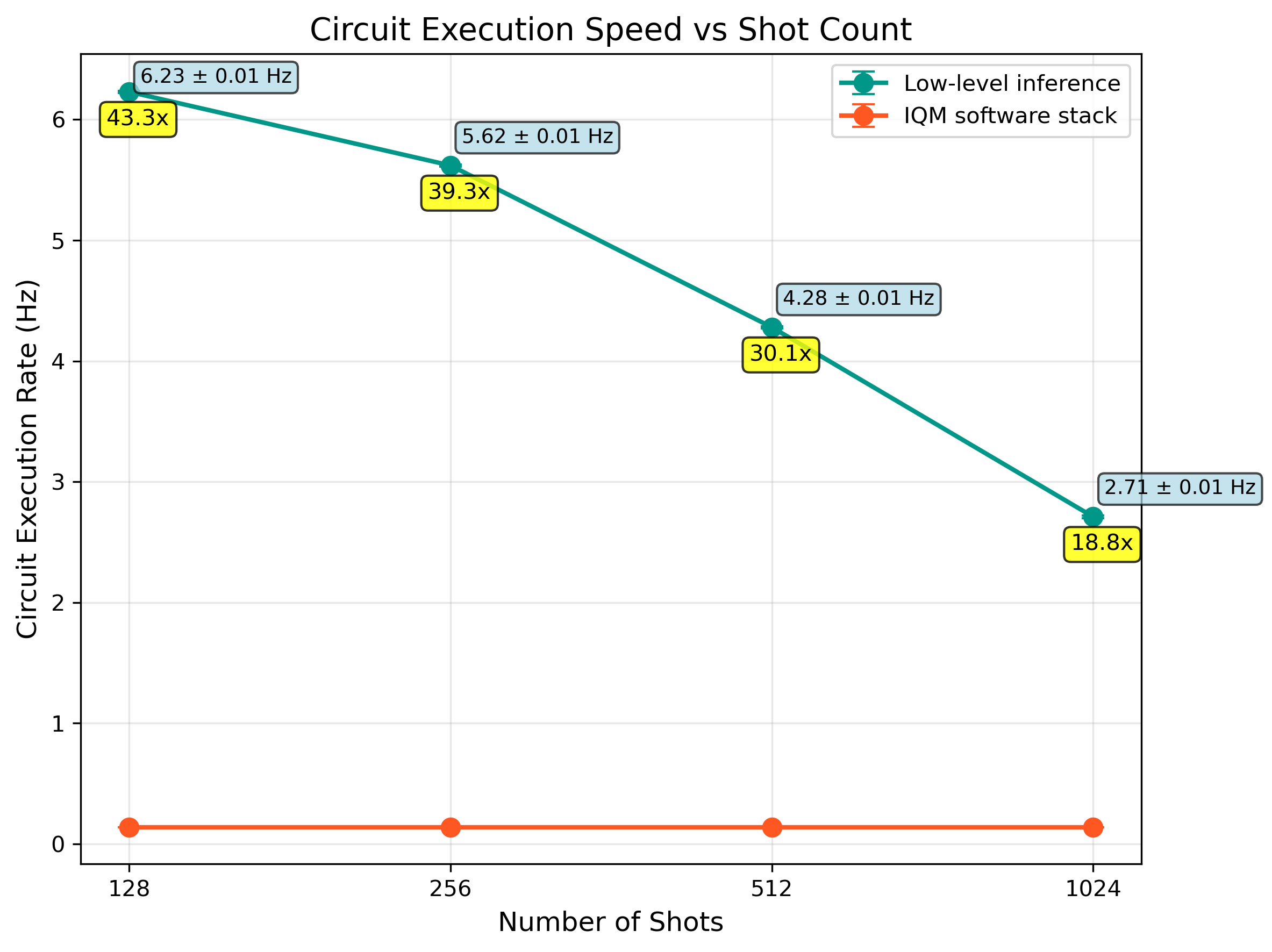}\\
    \small (a)
  \end{minipage}
  \hfill
  \begin{minipage}{0.48\textwidth}
    \centering
    \includegraphics[width=\textwidth]{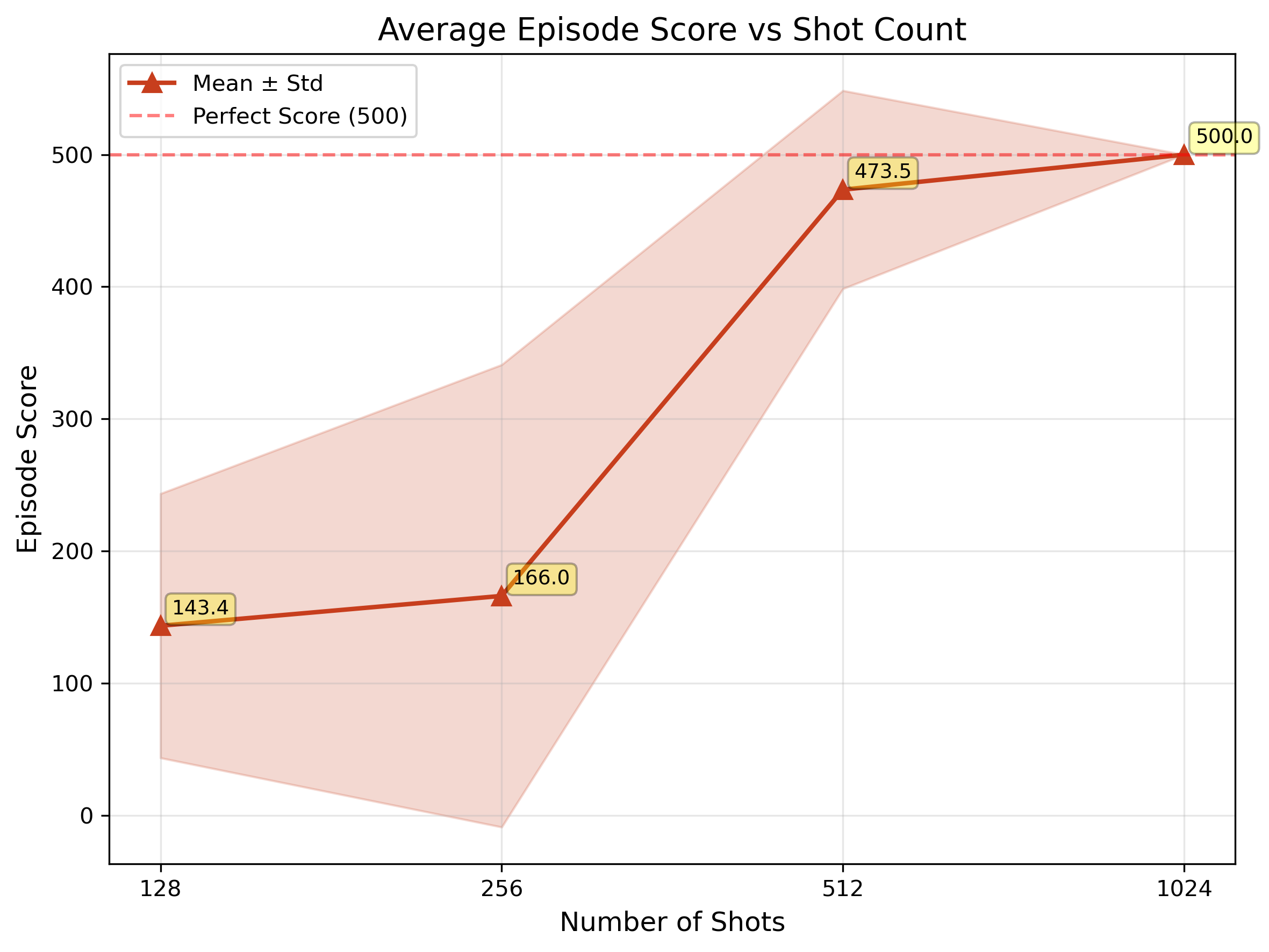}\\
    \small (b)
  \end{minipage}
  \caption{Average iteration rate and throughput (\textbf{a}) and corresponding inference-time episode scores (\textbf{b}) for shot counts of 128, 256, 512, and 1024. Error bars represent the standard deviation across five runs of ten episodes (\textbf{a}) and ten evaluation episodes (\textbf{b}). The red line at 500 points in panel (\textbf{b}) indicates perfect CartPole-v1 performance.}
  \label{fig:iteration_performance_scores}
\end{figure}

Figure~\ref{fig:iteration_performance_scores} summarizes the latency--performance trade-off. As expected, increasing the shot count reduces the iteration rate but improves control performance: average episode scores increase from $\approx$143 (128 shots) and $\approx$166 (256 shots) to $\approx$474 (512 shots), reaching perfect performance (500) at 1024 shots. These hardware runs use an inference control frequency of $f_{\text{c,inf}}=50$~Hz in a simulated, offline CartPole loop, and no readout error mitigation is applied. We emphasize that this operating point differs from the inference regimes identified as near-perfect in the duration-matrix study (which favored higher $f_{\text{c,inf}}$ and/or larger shot budgets under the \texttt{FakeAdonis} model); as such, the absolute episode scores here should be interpreted as a conservative hardware demonstration under non-ideal inference conditions. Together, the results show that low-level execution enables order-of-magnitude faster closed-loop inference on hardware while preserving high CartPole performance when the shot budget is sufficiently large in the case of no readout mitigation.

\begin{table}[!h]
\centering
\begin{tabular}{c|cc|c}
\hline
\textbf{Shots} & \textbf{IQM stack (iter/s)} & \textbf{Low-level (iter/s)} & \textbf{Speedup} \\
\hline
128   & 0.144 ($\pm$ 0.001) & 6.23 ($\pm$ 0.01) & \bm{$43.3\times$} \\
256   & 0.143 ($\pm$ 0.001) & 5.62 ($\pm$ 0.00) & \bm{$39.3\times$} \\
512  & 0.142 ($\pm$ 0.001) & 4.28 ($\pm$ 0.01) & \bm{$30.1\times$} \\
1024  & 0.144 ($\pm$ 0.002) & 2.71 ($\pm$ 0.01) & \bm{$18.8\times$} \\
\hline
\end{tabular}
\caption{Average iteration speeds for the IQM software stack and low-level inference at different shot counts, including computed speedups.}
\label{tab:iteration-speeds}
\end{table}

\iffalse
% Iteration Speed and Rate plots side by side
  \begin{figure}[t]
      \centering
      \includegraphics[width=0.7\textwidth]{Figures/manual_circuit_rate_plot_20251124_235054.png}
      \caption{Average iteration (a) and throughput (b) for the shot counts of 128, 256, 512, and 1024. Error bars represent standard deviation across five runs of ten episodes.}
      \label{fig:iteration_performance}
  \end{figure}

  % Average Episode Score as separate figure
  \begin{figure}[!h]
      \centering
      \includegraphics[width=0.7\textwidth]{Figures/iteration_speed_plot_20251124_110728_episode_scores.png}
      \caption{Average inference-time episode scores corresponding to the same shot-count configurations reported in Fig.~\ref{fig:iteration_performance}. The red line at 500 points indicates perfect CartPole-V1 performance. Error bars show the standard deviation across ten episodes.}
      \label{fig:episode_scores}
  \end{figure}

The episode scores shown in Fig.~\ref{fig:episode_scores} correspond to the very same runs used to obtain the iteration-time and throughput results in Fig.~\ref{fig:iteration_performance}. In other words, these scores were recorded from the episodes executed during the timing measurements for each shot-count configuration.
  \fi

  \FloatBarrier

\section{Discussion}

% We have demonstrated the successful control of a classical system by a quantum computer in an offline setting. The quantum agent learned more efficiently than the classical counterpart, which opens an interesting avenue to explore potential quantum speed-ups in learning control tasks. Furthermore, we identified a bottleneck when executing on real quantum systems, namely fast measurement and qubit reset, severely reducing the real-time capabilities of the system.  For the future, we plan to explore other hardware-software control stacks that might enable real-time execution \cite{Reuer-2023}. The restless measurement scheme \cite{Werninghaus-2021} is also a potential candidate ...
% Given that dynamic circuits are crucial for quantum error correction, we are optimistic that fast and precise feedback operations will be available in the future, unlocking the control of larger control systems.

%\iffalse
In this work we have combined model-level, simulational, and hardware-level investigations to assess whether a minimal hybrid quantum--classical agent can (i) learn CartPole faster than a standard classical baseline, (ii) map out how performance depends on mismatches between training and deployment conditions across quantum backends and control frequencies, and (iii) be executed fast enough on current NISQ hardware to approach real-time control.

On the learning side, our expectation–value simulations show that a single-qubit hybrid agent with the encoding and feature set used in this work can solve CartPole in substantially fewer episodes than a comparable classical actor–critic network (Figures~\ref{fig:score_comparison}). This is in line with results reported in another work \cite{Hsiao2022-uj,}. This advantage persists even when gradients are estimated via the parameter–shift rule, albeit with slightly slower convergence than in the fully analytical setting. Given that CartPole is a low-dimensional, well-understood benchmark, these results do not constitute a quantum speed-up in the complexity-theoretic sense, but they do indicate that parametrized single-qubit circuits can act as effective nonlinear function approximators within RL loops when paired with the chosen encoding and lightweight classical post-processing.

\iffalse
The training–inference compatibility study extends this picture to noisy and shot-based regimes. By scanning over training backends, control frequencies, and shot counts, we observe that performance on the \texttt{FakeAdonis} device model is highly sensitive to both the training configuration and the deployment frequency (Figures~\ref{fig:expval_inf_20_25Hz} and~\ref{fig:expval_inf_33_50Hz}). In particular, agents trained on an ideal expectation–value backend at an intermediate control frequency of $33$~Hz transfer most reliably across all tested inference frequencies on \texttt{FakeAdonis}, whereas agents trained at $20$~Hz fail to produce any ten-second balancing episodes under the same inference conditions. Together with the complementary shot-based compatibility experiments reported in Appendix~\ref{app:compatibility}, these findings indicate that naively matching training and inference conditions is not always optimal. Instead, there appears to be a ``sweet spot'' in the training configuration that balances optimization stability with good performance under device noise and realistic control-frequency constraints, echoing sim-to-real observations from classical RL as well as recent work on the dual role of noise in variational quantum optimization.
\fi

From the duration matrices in Figures~\ref{fig:fakeadonis_128_256shots} and~\ref{fig:fakeadonis_512_1024shots}, both the inference-time control frequency $f_{\text{c,inf}}$ and the inference shot budget $N_{\text{shots,inf}}$ have a strong, expected impact on the average balancing duration: faster closed-loop updates and more precise (higher-shot) expectation estimates generally improve stability and reduce variability. A particularly clear reference point is $f_{\text{c,inf}}=100$~Hz, where all tested shot counts (128--1024) achieve mean balancing durations very close to the 10~s episode limit and often reach the limit, largely independent of training frequency. Moving to lower inference frequencies, the shot count increasingly determines where stable control becomes feasible: increasing $N_{\text{shots,inf}}$ shifts the onset of near-perfect performance to lower $f_{\text{c,inf}}$ (i.e., configurations that require 100~Hz at low shot budgets can succeed already around 50~Hz, and in some cases 33~Hz, when the shot count is increased). Notably, these results are obtained without applying readout error mitigation, so improved performance—especially in the lower-frequency / lower-shot regimes—would be expected with standard mitigation techniques. Finally, across the full sweep we observe no consistent dependence on the training control frequency $f_{\text{c,train}}$: for a fixed inference configuration $(f_{\text{c,inf}}, N_{\text{shots,inf}})$ the inference balancing durations remain broadly similar across all $f_{\text{c,train}}$, suggesting that deployment is governed primarily by inference-time constraints rather than the control frequency used during training.

Our hardware experiments on VTT Q5 address the practical question of whether such learned quantum policies can, in principle, be executed quickly enough for feedback control. By bypassing the high-level quantum control stack and directly programming the Zurich Instruments AWG and readout electronics via command tables, we achieve more than an order-of-magnitude improvement in iteration rate compared to the standard IQM stack (Table~\ref{tab:iteration-speeds}). The side-by-side timing and performance results in Figure~\ref{fig:iteration_performance_scores} show that this low-level execution path maintains high CartPole scores while substantially reducing latency, yielding iteration rates in the range of 2.7--6.2~Hz depending on the shot count. Future work could explore more advanced sampling schemes. For example, active reset after the final measurement can reduce the inter-shot wait time significantly. Another simpler approach is the restless measurement scheme \cite{Werninghaus-2021} where qubits are not reset between shots. However, the drawback of this approach is the requirement to make the circuit invariant with respect to the initial state being either $|0\rangle$ or $|1\rangle$. Finally, the control electronics could be updated to the latest generation of devices or replace by custom made instruments which support more granular control of drive and readout and reduced latencies. 

Some limitations of our study point to natural directions for future work. First, all experiments are conducted on a single-qubit architecture without entanglement; extending the analysis to few-qubit ansätze would allow us to probe how the observed learning-speed trends and compatibility patterns scale with circuit size and the number of trainable parameters. In addition, one could explore ansätze whose hardware-level implementation would be independent of the discrete initial state ($|0\rangle$ or $|1\rangle$), so that the state would not necessarily require a reset after each shot. Such designs could offer further speedups, but they impose nontrivial structural constraints on the circuit, are very likely to reduce degrees of freedom within the state, and potentially become increasingly challenging to realize as the number of qubits grows.

On the hardware side, our results suggest concrete performance targets for quantum devices aimed at control applications. Closing the remaining gap between our current multi-hertz iteration rates and the tens-of-hertz regime required for real-time feedback would enable not only CartPole control in the lab, but also more ambitious demonstrations.

\section{Conclusion}

This work addresses the gap between theoretical quantum reinforcement learning and physical deployment by presenting an end-to-end investigation of a hybrid agent on the CartPole benchmark. We demonstrated that a minimal single-qubit model acts as a sample-efficient policy, solving the environment in significantly fewer episodes than a classical baseline even when trained under realistic parameter-shift gradients. Evidently, our hardware experiments quantified the latency bottlenecks inherent to standard cloud-access models. By bypassing the high-level software stack and programming the control electronics directly, we achieved inference rates of over 6 Hz—a speedup of more than 40x compared to standard execution. These results, combined with our analysis of the trade-offs between shot budget and control frequency, outline a roadmap for turning small-scale quantum RL demonstrations into practical, closed-loop quantum controllers where latency and precision must be strictly balanced.

\printbibliography

%\subsubsection{Subsubsection heading}
%Sample text inserted for demonstration.

% \begin{figure}
%  \centering
%         \includegraphics[width=0.5\textwidth]{figure1}
%  \caption{Text describing the figure and the main conclusions drawn from it. To make your figures accessible to as many readers as possible, try to avoid using colour as the only means of conveying information. For example, in charts and graphs use different line styles and symbols. Further information is available in the online guide: \href{https://publishingsupport.iopscience.iop.org/publishing-support/authors/authoring-for-journals/writing-journal-article/\#figures}{https://publishingsupport.iopscience.iop.org/publishing-support/authors/authoring-for-journals/writing-journal-article/\#figures}}
% \label{fig1}
% \end{figure}

%
% Each of the commands below will create an unnumbered section with the appropriate heading.
% Remove any sections that are not relevant for your article.
% All sections except suppdata will be removed if the [anonymous] option is used.
% See iopjournal-guidelines.pdf for more information.
%

\ack{J.Q.Q. and P.W. acknowledges the financial support of the Quantum Technology Future Science Platform - CSIRO.}

%\funding{Sample text inserted for demonstration.}
% This section is a list of funder names and grant numbers

% List author names and the contributions made to the article, using terms from the NISO Contributor Roles Taxonomy (CRediT) https://credit.niso.org
\roles{

N.T.T.N.: Conceptualization, Methodology (lead on hybrid quantum-classical architecture and single-qubit algorithm design), Software (development of RL agents and simulation environment), Formal analysis (simulation and compatibility studies between QRL and CRL models), Writing – original draft.

\noindent{V.M: Conceptualization, Methodology (experiments on the training frequency and inference frequency), Software (experiments on inference), Formal analysis (simulation and compatibility studies between QRL and CRL models), Writing – original draft.}

\noindent{J.J: Methodology (hardware setup on VTT Q5), Software (VTT Q5 control stack), Formal analysis (circuit execution speed and shot count), Writing – original draft.}

\noindent{T.-F.L.: Supervision, Resources, Writing – review $\&$ editing. J.Q.Q.: Conceptualization, Supervision, Project administration, Funding acquisition, Writing – review $\&$ editing.}
}

%10.5281/zenodo.17720358
\data{The data supporting the plots can be found at 10.5281/zenodo.18994218. Source code and model weights are available upon reasonable request.}
% For more information on IOP Publishing's research data policy see: https://publishingsupport.iopscience.iop.org/questions/research-data/

\newpage
\appendix

\section{Gradient Computation}
\label{subsec:gradients}
%rumelhart1986learning
Backpropagation is a fundamental technique in training machine learning models, in which trainable parameters are iteratively updated to minimize a loss function~\cite{rumelhart1986learning}. When training on quantum simulators—such as Qiskit \cite{javadi2024quantum} or PennyLane \cite{bergholm2018pennylane}—this process is automated via seamless integration with PyTorch's automatic differentiation library \cite{paszke2019pytorch}. However, transitioning to physical quantum hardware disrupts this workflow. The Quantum Processing Unit (QPU) operates as an external execution environment outside the scope of the classical computational graph. Because the internal operations on the remote QPU are not tracked by PyTorch, and the quantum states collapse upon measurement, automatic gradient computation is rendered impossible.

Consequently, when training on a Noisy Intermediate-Scale Quantum (NISQ) device, the quantum parameters $\theta$ are effectively decoupled from the classical automatic differentiation engine. To address this, the gradients for the quantum variational circuit must be calculated separately and bridged with the classical neural network. During the backward propagation step, the gradient of the loss function with respect to the trainable parameters $\theta$, denoted as $\nabla L(\theta)$, is required for the optimization algorithm (e.g., Adam). By applying the chain rule, this gradient can be expressed as:
\begin{equation}
    \nabla L(\theta) = \frac{\partial L_{actor}}{\partial \theta} = \frac{\partial L_{actor}}{\partial f(s_i, \theta)} \cdot \frac{\partial f(s_i, \theta)}{\partial \theta}
\end{equation}
Here, the term $\frac{\partial L_{actor}}{\partial f(s_i, \theta)}$ represents the gradient of the loss function with respect to the quantum circuit's output. Since $f(s_i, \theta)$ serves as the input to the subsequent classical layers, this term can be computed using standard backpropagation via PyTorch.
The second term, $\frac{\partial f(s_i, \theta)}{\partial \theta}$, represents the gradient of the quantum circuit output with respect to its own parameters. This is typically calculated using the parameter-shift rule \cite{wierichs2022general}:
\begin{equation}
    \frac{\partial f(s_i, \theta)}{\partial \theta} = \frac{1}{2} \left[ f\left(s_i, \theta + \frac{\pi}{2}\right) - f\left(s_i, \theta - \frac{\pi}{2}\right) \right]
\end{equation}
This approach allows for the analytical calculation of gradients by executing additional circuit evaluations with shifted parameters on the physical hardware. These values are then fed back into the classical optimizer to update the model weights.

\section{Additional Experiment Details}
\label{app:compatibility}

\begin{table}[h]
\centering
\begin{tabular}{lc}
\hline
\textbf{Parameter} & \textbf{Value} \\
\hline
Drive $\omega_d / 2\pi$ (GHz)   & 4.211 \\
Drive duration (ns) & 120 \\
Readout frequency $\omega_{RO}/2\pi$ (GHz) & 6.157 \\
Readout duration ($\mu$s) & 1  \\
Reset wait between shots ($\mu$s) & 398 \\
\hline
$T_1$-time ($\mu$s) & $22.94 \pm 2.03$ \\
$T_2^*$-time ($\mu$s) & $15.68 \pm 2.09$ \\
$T_2$-time ($\mu$s) & $26.19 \pm 4.92$ \\
Single-qubit RB (\%) & $99.76 \pm 0.01$ \\
Single-shot readout fidelity (\%) & 95.45 \\
Single-shot readout 0-1 error (\%) & 2.95 \\
Single-shot readout 0-1 error (\%) & 6.15 \\
\hline
\end{tabular}
\caption{Qubit parameters, coherence times and gate fidelities as well as circuit execution details.}
\label{tab:hardware-details}
\end{table}

\FloatBarrier
\begin{figure}[htbp]
  \centering
  \includegraphics[width=0.7\textwidth]{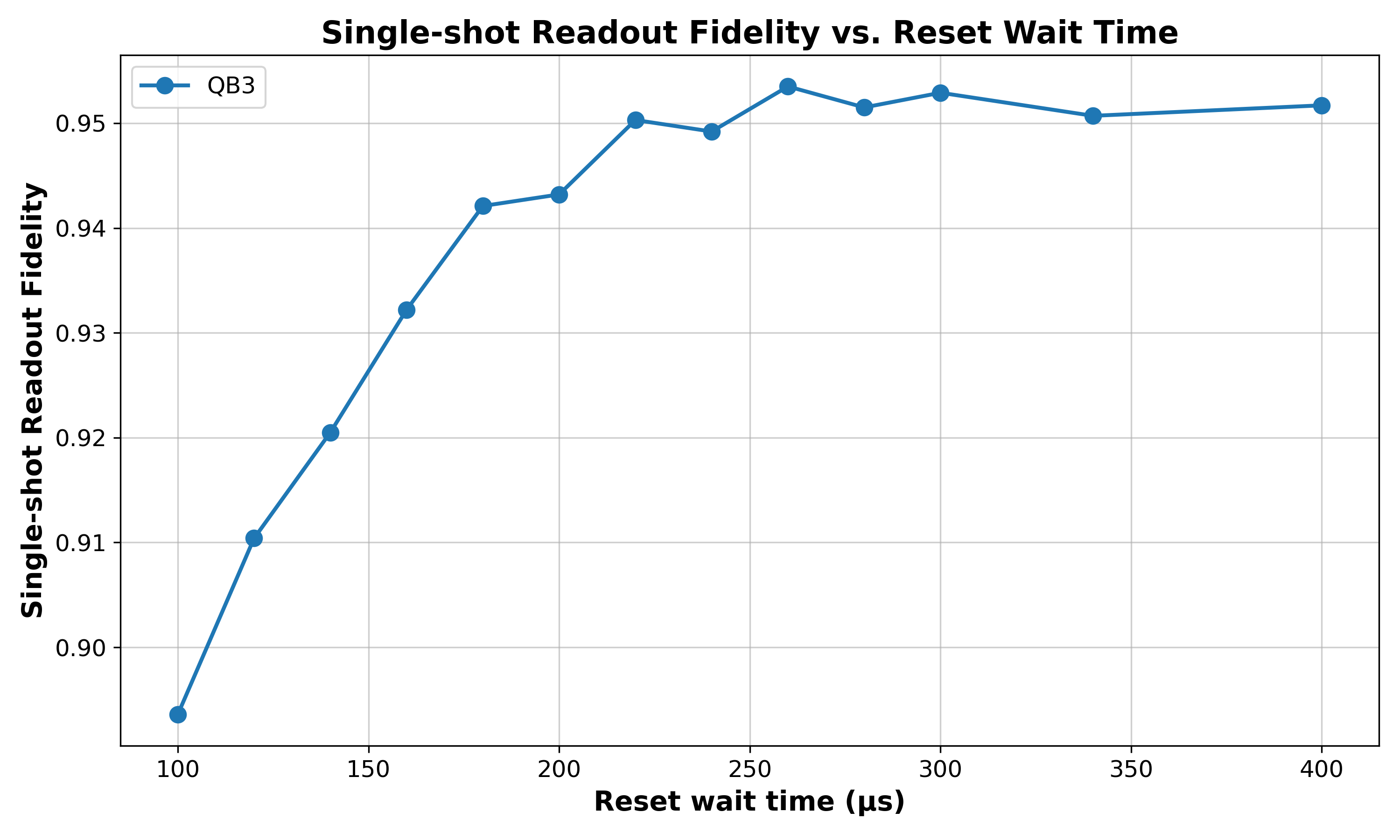}
  \caption{Single-shot readout fidelity as a function of reset wait time ($\mu$s) for qubit 3 on VTT Q5. The reset wait time is the time between two shots to ensure that the qubit relaxes to the ground state. For each wait time 40.000 shots were executed.}
  \label{fig:sweep_reset_wait_time}
\end{figure}

\end{document}